\documentclass[aps,prl,twocolumn,showpacs,superscriptaddress,groupedaddress]{revtex4}  
\usepackage{graphicx}  
\usepackage{dcolumn}   

\usepackage{bm}        
\usepackage{amssymb}   
\usepackage{amsmath}
\usepackage{threeparttable}
\usepackage{tablefootnote}

\usepackage[caption=false]{subfig}
\usepackage{verbatim}
\usepackage{multirow}

\def\fiverm{\footnotesize}

\usepackage[usenames,dvipsnames,svgnames]{xcolor}

\def\2;{\;\;}

\def\eps{\epsilon}

\def\pr{{\prime}}

\def\IntZ{{\mathbb Z}}

\def\w#1{\widetilde{#1}}
\def\Ref#1{(\ref{#1})}

\def\C#1{{\mathcal #1}}

\def\binom#1#2{{{#1}\choose{#2}}}
\def\Bi#1#2{{\binom{#1}{#2}}}

\def\Sfrac#1#2{\hbox{\large $\frac{#1}{#2}$}}

\def\svv{{\;\hbox{$|$}\;}}

\def\plus{{\hspace{0.85pt}{+}\hspace{0.85pt}}}
\def\minus{{\hspace{0.85pt}{-}\hspace{0.85pt}}}

\begin{document}

\widetext
\leftline{Version 1.0 as of \today}

\title[Entropic exponents of grafted lattice stars]{
Entropic exponents of grafted lattice stars}
\author{EJ Janse van Rensburg}
\affiliation{Department of Mathematics and Statistics, 
York University, Toronto, Ontario M3J~1P3, Canada 
(Email: rensburg@yorku.ca)\\}

\date{\today}

\begin{abstract}
The surface entropic exponents of half-space lattice stars grafted at their central 
nodes in a hard wall are estimated numerically using the PERM algorithm.
In the square half-lattice the exact values of the exponents are verified, 
including Barber's scaling relation and a generalisation for $2$-stars
with one and two surface loops respectively.   This is the relation
\[ \gamma_{211}=2\,\gamma_{21}-\gamma_{20},\]
where $\gamma_{21}$ and $\gamma_{211}$ are the surface entropic exponents 
of a grafted $2$-star with one and two surface loops respectively, and 
$\gamma_{20}$ is the surface entropic exponent with no surface loops.
This relation is also tested in the cubic half-lattice where surface entropic 
exponents are estimated up to $5$-stars, including many with one 
or more surface loops. Barber's scaling relation and the relation 
\[ \gamma_{3111}=\gamma_{30}-3\,\gamma_{31}+3\,\gamma_{311} \]
are also tested, where the exponents $\{\gamma_{31},\gamma_{311},\gamma_{3111}\}$ 
are of grafted $3$-stars with one, two or three surface loops respectively,
and $\gamma_{30}$ is the surface exponent of grafted $3$-stars.
\end{abstract}

\maketitle

\section{Introduction}

The \textit{connectivity} of a polymer network can be represented as an
abstract graph $\C{G}$ of nodes and bonds, where the nodes are branching
points in the network, and the bonds represent linear polymers joining the nodes
into the network.  If the network is embedded in a lattice, then it is a model
of the connectivity and the topology of a polymer network in the plane 
or in a thin layer (if the lattice is two dimensional), or in a good solvent
(if the lattice is three dimensional).  

If a network with connectivity $\C{G}$ is embedded in the (hyper)-cubic lattice 
$\IntZ^d$, then the nodes are located on vertices in the lattice, and the bonds are 
mutual- and self-avoiding walks joining the nodes.  These self-avoiding walks are the
\textit{branches} of the network.  If $\C{G}$ is a star graph, then the branches are
called \textit{arms}.  The embedding is \textit{uniform} or \textit{monodispersed}
if all the branches are walks of the same length.  

The lattice embedding is a model that quantifies the entropy of the corresponding
polymer network.  If $c_n(\C{G})$ is the number of distinct embeddings of the network, 
counted up to translation (or by fixing a node at the origin), then the usual scaling 
assumption is
\begin{equation}
c_n(\C{G}) \sim n^{\gamma(\C{G})-1}\mu_d^n
\label{1}
\end{equation}
where $\gamma(\C{G})$ is the \textit{entropic exponent} of the network
and $\mu_d$ is the self-avoiding walk growth constant \cite{HM54,BH57}
(see references \cite{SS93,WS92}, and for lattice stars reference \cite{CW87}).
The best estimates of the growth constants in the square and cubic lattices
are obtained from simulations of the self-avoiding walk, and are
\begin{align}
\mu_2 &= 2.63815853035(2),  \;\hbox{\cite{CJ12}}
\label{1b} \\
\mu_3 &= 4.684039931(27).  \;\hbox{\cite{C13}}
\label{1c}
\end{align}

Uniform lattice star polymers form a particular class of lattice networks that have
received significant attention in the literature at least since the 1980s
\cite{MF84,LWWMG85,CW87,BK89,WS92,SS93,DG20}.   If $s_n^{(f)}$ is the number
of uniform lattice stars in $\IntZ^d$ with $f$ arms (these are \textit{$f$-stars}) with
central node at the origin, then by equation \Ref{1},
\begin{equation}
s_n^{(f)} \sim n^{\gamma_f -1}\,\mu_d^n ,
\label{2}
\end{equation}
where $\gamma_f$ is the $f$-star entropic exponent. High quality numerical 
results and estimates of the entropic exponents of star polymers were made in references 
\cite{OB91,GG94,O02,HNG04,CJvR21,CJvR21a}.

\begin{figure}[t!]
\includegraphics[width=0.45\textwidth]{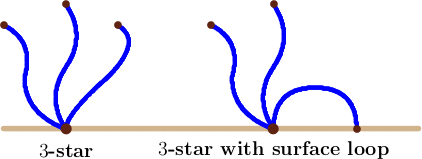}
\caption{\textit{Schematic of $3$-stars in a half-lattice.  The central node of the
star is attached at the origin in the hard wall (the boundary of the half-lattice).  On 
the right the star has one arm forming a surface loop.  The star on the left has no
loops or surface loops.}}
\label{f1}  
\end{figure}

Lattice networks of connectivity $\C{G}$ grafted to a wall have scaling similar
to equation \Ref{1}.  In the hypercubic half-lattice $\IntZ^d_+ = 
\{ (x_1,x_2,\ldots,x_d)\in\IntZ^d \svv \hbox{such that $x_d\geq 0$} \}$
a chosen node of a network is grafted at the origin.  Nodes of the network located 
in the boundary (or hard wall) $\partial \IntZ^d_+$ of the half-lattice are
\textit{surface nodes}. Depending on the connectivity $\C{G}$ such
networks may have \textit{loops} (circuits which are lattice polygons) as well
as \textit{surface loops} (self-avoiding walks with both endpoints in 
hard wall $\partial \IntZ^d_+$ of the half-lattice).  For example, in figure
\ref{f1} we show a schematic of a lattice network in $\IntZ^2_+$ with 
connectivity a lattice $3$-star without (left) and with (right) a surface loop.

\subsection{Vertex exponents, and entropic exponents}

The entropic exponents of networks of connectivity $\C{G}$, in the full
lattice, or in the half-lattice, can be expressed in term of \textit{vertex exponents} 
$\sigma_f$ and \textit{surface vertex exponents} $\sigma^\prime_g$ by 
\cite{D86,D89C}
\begin{equation}
\gamma_\C{G} = 1+ \sum_f m_f\sigma_f + \sum_g m_g \sigma^\prime_g
-c(\C{G})\,dv - \ell(\C{G})\,\nu
\label{3}
\end{equation}
where $\nu$ is the metric exponent of the self-avoiding walk ($\nu=3/4$
in two dimensions \cite{N84}, and $\nu=0.58759700(40)$ in three dimensions 
\cite{CD16}).  The coefficient $m_g$ is the number of surface-nodes of degree 
$g$ in $\partial\IntZ^d_+$, $m_f$ is the number of nodes of degree $f$ in the bulk
lattice, $c(\C{G})$ is the number of independent circuits in the network, and 
$\ell(\C{G})$ is the number of independent surface loops. 
Testing of equation \Ref{3} are limited to self-avoiding walks in a half-lattice 
\cite{G04}, and for branched networks, to star polymers in bulk 
\cite{HNG04,CJvR21,CJvR21a} and a few cases of branched acyclic networks
in bulk \cite{CJvR21,CJvR21a}.  

Exact values of the vertex exponents in two dimensions were calculated using 
conformal invariance techniques \cite{D86,D89C,DS86}.  These are
\begin{equation}
\sigma_f = \Sfrac{1}{64}\,(2{-}f)\,(9f{+}2),
\quad\hbox{and}\quad 
\sigma_f^\prime = \Sfrac{1}{32}\,f\,(2{-}9f) .
\label{6a}
\end{equation}
Using these expressions, the exact values of $\gamma_{\C{G}}$ are known in
the square lattice. For example, putting $f=0$ and using equation \Ref{3}
give the exact value of the entropic exponent of the self-avoiding walk
$\gamma=1+2\sigma_1=43/32$ \cite{N84}.

In three dimensions there are $\epsilon$-expansion estimates \cite{GZJ98}
for the vertex exponents.  To first order in $\epsilon$ \cite{D86}, 
\begin{align}
\sigma_f & = \Sfrac{\epsilon}{16}\,f\,(2{-}f) + O(\epsilon^2);
\label{7a} \\
\sigma_f^\prime &= - \Sfrac{1}{2}\,f + \Sfrac{\epsilon}{16}\,f\,(2{-}f) + O(\epsilon^2) .
\label{8a}
\end{align}
These approximations deteriorate quickly with increasing degrees of the nodes
(see for example \cite{CJvR21a}), but can be used with equation \Ref{3}
to approximate the entropic exponents $\gamma_{\C{G}}$ in three dimensions.
In the case of the self-avoiding walk, the first order $\epsilon$-expansion estimate
$\gamma=1{+}2\sigma_1\approx1.125$ compares relatively well with the best 
numerical estimates $\gamma=1.15698(34)$ \cite{SBB11} and
$\gamma=1.156 953 00(95)$ \cite{C17}. See references \cite{D86,D89C,SFLD92}
for $O(\eps^2)$, and \cite{SFHFB03} for $O(\eps^4)$, expansions for $\sigma_f$.

\begin{figure}[t!]
\includegraphics[width=0.45\textwidth]{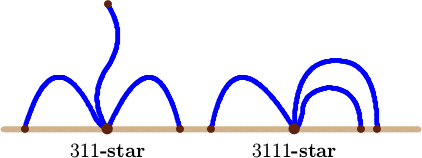}
\caption{\textit{Schematic illustrations of $311$-stars (left) and
$3111$-stars (right) in a half-lattice.  In the half-square lattice the
one arm of the $311$-star is ``screened'' from the hard wall by
the other two arms with endpoints in the hard wall.  In the
case of $3111$-stars all three arms have endpoints in the
hard wall, and the only way to accommodate this is by having
one arms located inside the surface loop created by the hard wall and an
arm.  This makes simulations of $3111$-stars in the square half-lattice
very challenging.  In our PERM simulations we rarely observed these
states in the $10^9$ iterations done.}}
\label{figure2}  
\end{figure}

In this paper the focus is on a lattice star in the half-lattice with its central node 
at the origin, such as schematically illustrated in figure \ref{f1}.  We first introduce
notation to distinguish the connectivities of $f$-stars in the half-lattice
$\IntZ^d_+$ efficiently.   

Let $f1^g \equiv f\overbrace{11\ldots1}^{\hbox{$g$}}$ denote lattice stars in the
half-lattice $\IntZ^d_+$ with central node of degree $f$ located at
the origin in the boundary $\partial\IntZ^d_+$ (\textit{the hard wall}), and
with $g\leq f$ arms having their endpoints in the hard wall (and so forming 
$g$ \textit{surface loops}).  For example, in figure \ref{f1} the  cases
$31^0 \equiv 30$ (left) and $31^1\equiv 31$ (right) are shown.  Similarly,
$31^2 \equiv 311$ denotes $3$-stars with central node at the origin and
with $2$ arms having their end-points in the hard wall (and so forming
two surface loops), and $31^3\equiv 3111$ denotes $3$-stars with three surface 
loops (see figure \ref{figure2}).

The entropic exponents of half-lattice stars with connectivity $f1^g$ is
denoted by $\gamma_{f1^g}$.  For example, $\gamma_{311}$ is the
entropic exponent of $3$-stars with two arms forming surface loops,
and for the cases in figure \ref{1}, we have $\gamma_{30}$ (left) and 
$\gamma_{31}$ (right).  In terms of equation \Ref{3} the entropic
exponent of $f$-stars with $g$ surface loops in the half-lattice is given by
\begin{equation}
\gamma_{f1^g}= \gamma_{f\underbrace{11\ldots1}_{\hbox{$g$}}}
= 1+ \sigma_f^\prime + (f{-}g)\,\sigma_1 + g\,\sigma_1^\prime - g\,\nu .
\label{4}
\end{equation}
The vertex exponents are given in equation \Ref{6a} in two dimensions,
and are approximated in the $\epsilon$-expansion in equations \Ref{7a} 
and \Ref{8a} in three dimensions.

In the case of a self-avoiding walk $\gamma=1{+}2\sigma_1$, and for
a walk from the origin in $\IntZ^d_+$, $\gamma_1
=1{+}\sigma^\prime_1{+}\sigma_1$ (since $f=1$ and $g=0$).  
If both endpoints are in $\partial\IntZ^d_+$, then $f=g=1$ and thus 
$\gamma_{11}=1{+}2\sigma^\prime_1-\nu$.
Eliminating $\sigma_1$ and $\sigma^\prime_1$ gives Barber's
scaling relation \cite{B73,BGM78}
\begin{equation}
2\,\gamma_1-\gamma_{11} = \gamma + \nu .
\label{10}
\end{equation}
In $d=2$ the sum of the exact values $\gamma=43/32$, $\nu=3/4$ \cite{N84}
on the right hand side equals the sum of the exact values $\gamma_1=61/64$ 
\cite{C87} and $\gamma_{11}=-3/16$ \cite{DS86} on the left hand side.
Numerical results are in agreement with these results, namely
$\gamma_1=0.945(5)$ and $\gamma_{11}=-0.19(3)$ \cite{BGM78}, and 
$\gamma_{1}=0.9551(3)$ \cite{MC93}.  This shows that $2\,\gamma_1
-\gamma_{11}=2.08(4)$ while the exact value is $\gamma{+}\nu
=67/32=2.09375\ldots$. 

In three dimensions early estimates $\gamma_1=0.687(5)$ and 
$\gamma_{11}=-0.38(2)$ \cite{LM88A,ML88A}, and $\gamma_1=0.697(2)$ 
and $\gamma_{11}=-0.383(5)$ \cite{HG94},  gave way to the more accurate 
results $\gamma_1=0.6786(12)$ and $\gamma_{11}=-0.390(2)$ in 
reference \cite{G04}.  The best available estimates can be found in reference
\cite{CCG16} where $\gamma_1$ is estimated, and $\gamma_{11}$ is estimated
from the bridge entropic exponent $\gamma_b$ using the relation
$\gamma_b = \gamma_{11} + \nu$ \cite{DG20}.  These are
$\gamma_1= 0.677667(17)$ and $\gamma_{11}=-0.389245(28)$.
These values give $2\gamma_1{-}\gamma_{11}=1.744579(62)$
and using the best estimates $\gamma=1.156 953 00(95)$ \cite{C17} and
$\nu=0.58759700(40)$ \cite{C10,CD16}, $\gamma{+}\nu = 1.7445500(14)$.  This
confirms the Barber scaling relation to very high accuracy.
The mean field values of these exponents are $\gamma=1$, $\nu=1/2$,
$\gamma_1=1/2$ and $\gamma_{11}=-1/2$ \cite{BDG83,BL90}.

More generally, one may notice that when $f\geq 2$, then the vertex 
exponents in equation \Ref{4} can be eliminated by using an alternating sum 
and binomial coefficients, so that
\begin{equation}
\sum_{g=0}^f (-1)^g \Bi{f}{g}\,\gamma_{f1^g} = 0.
\label{6}
\end{equation}
This shows, for example, that $\gamma_{20}-2\,\gamma_{21}+\gamma_{211} = 0$
and $\gamma_{30}-3\,\gamma_{31}+3\,\gamma_{311}-\gamma_{3111}=0$
and so on.   The identity
\begin{equation}
\gamma_{20}=\gamma-1
\label{12x}
\end{equation}
was noted in reference \cite{D86}) and from it and equation \Ref{6}
with $f=2$ it follows that 
\begin{equation}
2\,\gamma_{21}-\gamma_{211} = \gamma - 1.
\label{12}
\end{equation}
This is a generalisation of Barber's scaling relation (equation \Ref{10}).
Noting that $\gamma\minus 1 = 2\sigma_1$, $\gamma_1\minus 1
= \sigma_1+\sigma_1^\pr$, and $\gamma_{20}=\gamma\minus 1
= 1 \plus \sigma_2^\prime\plus 2\sigma_1$ (see also reference \cite{D89C}) 
allows one to solve for $\{\sigma_1,\sigma_1^\prime,\sigma_2^\prime\}$
from the best numerical estimates in references \cite{C17,DG20} to obtain
very accurate estimates for grafted $2$-star exponents:
\begin{eqnarray}
\gamma_{20} &=& 0.15695300(95) \nonumber \\
\gamma_{21} &=& -0.909930(17) \\ 
\gamma_{211} &=& -1.976813(33) \nonumber
\label{e14}
\end{eqnarray}
There estimates are consistent with equation \Ref{6} for $f=2$.

\section{Numerical results}

The numerical approaches developed in reference \cite{G04} based on the 
PERM algorithm \cite{G97,HNG04,HG11,CJvR21,CJvR21a}, and in particular 
the flat histogram \cite{PK04} and the parallel implementations \cite{CJvR20}
of PERM, can be used to estimate lattice star entropic exponents in $\IntZ^d$
(and in $\IntZ^d_+$) efficiently (see reference \cite{G04} for half-space 
self-avoiding walk sampling using PERM).  In this paper similar approaches are used, 
except that, in addition to self-avoiding walks grafted at one endpoint in 
$\partial\IntZ^d_+$, $f$-stars are sampled with their central nodes at the
origin in $\IntZ^d_+$.  The details of our simulations are shown in table \ref{t1}.
An iteration is a started PERM sequence (which may be pruned and enriched by
the algorithm). In these similutions the mersenne twistor random number 
generator \cite{MN98} was used, except in one case as noted in table \ref{t1}, 
where the Panneton generator in reference \cite{PLM06} was used instead.

\def\w{\textcolor{white}{$-$}}
\begin{table}[h!]
\caption{PERM simulations}
\begin{tabular}{@{}*{3}{l}ccc}
Dimension & Star & Length & Threads & Iter/Thread & Iterations \cr
\hline
\vspace{-2mm}
& & & & & \cr
\multirow{3}{*}{$d=2$} & $1$ & $10000$ & $4$ & $2.5\times 10^8$ & $10^9$ \cr
          & $2$ & $16000$ & $4$ & $2.5\times 10^8$ & $10^9$ \cr
          & $3$ & $15900$ & $4$ & $2.5\times 10^8$ & $10^9$ \cr
\hline
\vspace{-2mm}
& & & & & \cr
\multirow{5}{*}{$d=3$} & $1$ & $10000$ &  $4$ & $2.5\times 10^8$ & $10^9$ \cr
          & $1^*$ & $10000$ &  $8$ & $1.25\times 10^8$ & $4.04\times 10^8$ \cr
          & $2$ & $10000$ & $4$ & $2.5\times 10^8$ & $10^9$ \cr
          & $3$ & $9900$   & $6$ & $1.666\times 10^8$ & $10^9$ \cr
          & $4$ & $10000$ & $4$ & $2.5\times 10^8$ & $10^9$ \cr
          & $5$ & $12500$ & $4$ & $2.5\times 10^8$ & $10^9$ \cr
\hline
\multicolumn{3}{l}{$*$ -- Panneton generator \cite{PLM06}} \\
\end{tabular}
\label{t1}
\end{table}

In each simulation the data were sieved by collecting data on stars separately
by number surface loops.  Thus, the algorithm produced 
data on $s_n^{(f)}(g)$, the number of stars with central node of
degree $f$ at the origin in the half-lattice $\IntZ^d_+$, and with
$g$ arms forming surface loops.  The scaling of $s_n^{(f)}(g)$ is
\begin{equation}
s_n^{(f)}(g) \sim n^{\gamma_{f1^g}-1}\,\mu_d^n\,(1+B/n + C/n^\Delta + \cdots) ,
\label{9}
\end{equation}
where $\Delta$ is the (first) self-avoiding walk confluent correction exponent which has 
value $\Delta=3/2$ if $d=2$ \cite{N82,N84,CGJPRS05} and $\Delta=0.528(8)$ 
if $d=3$ \cite{CD16}. Dividing by $\mu_d^n n^{\gamma_{f1^g}-1}$ 
and taking logarithms give
\begin{equation}
\scalebox{0.96}{$\displaystyle
\log\left( \frac{s_n^{(f)}(g)}{\mu_d^n\,n^{\gamma_{f1^g}-1}} \right) =  
\begin{cases}
A + B/n + \cdots, & \hbox{if $d=2$}; \cr
 &
\label{8} \\
A + B/n^{\Delta} + \cdots, & \hbox{if $d=3$},
\end{cases}$}
\end{equation}
where the logarithms on the right hand side were expanded assuming $n$ is large.
The best value of $\gamma_{f1^g}$ and a confidence interval on it can be 
determined by plotting the left hand side against $1/n$ if $d=2$, and against 
$1/n^{\Delta}$ if $d=3$.  This approach was developed in reference \cite{G04} 
where it was used effectively for estimating $\gamma_1$ and $\gamma_{11}$
using PERM simulations in the cubic half-lattice.  See also references \cite{HNG04,CJvR21,CJvR21a}. 

\begin{figure}
\includegraphics[width=0.475\textwidth]{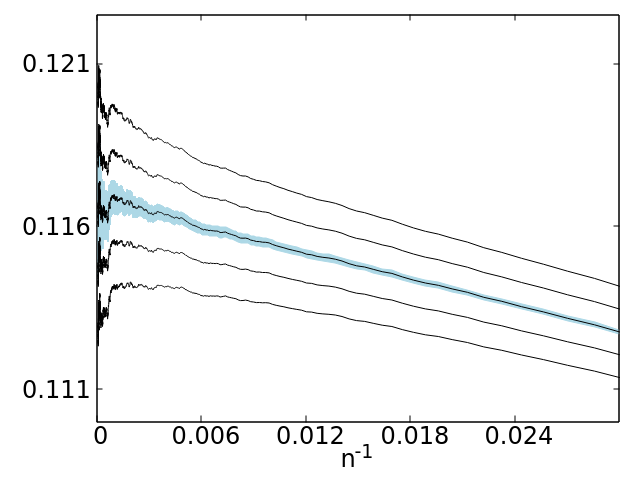} \\
\vspace{3mm}
\includegraphics[width=0.475\textwidth]{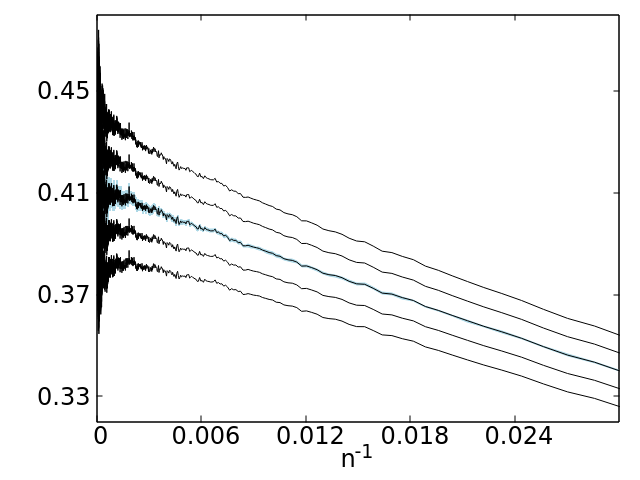}
\caption{\textit{Determining $\gamma_1$ and $\gamma_{11}$ in the square lattice
by plotting the left-hand side of equation \Ref{8} as a function of $1/n$.  The middle 
curve is obtained by selecting that value of the exponent giving a straight line.  
In each panel a shaded envelope (more visible in the top panel) on the middle curve
gives the confidence interval on the raw data.
(Top panel) The middle curve is plotted with $\gamma_1=0.95325$ against
$1/n$.  A confidence interval $\sigma=0.00020$ is obtained by adding and
subtracting $\sigma$ from $\gamma_1$.  The top two curves have an increasing
upwards tendency as the $y$-axis is approached, and the bottom two curves 
similarly have a downwards tendency.  This gives the best (rounded) estimate
$\gamma_1=0.9533\pm 0.0002$. (Bottom panel) A similar analysis to determine
$\gamma_{11}$.  This gives $\gamma_{11}=-0.188\pm0.002$.}}
\label{f2}
\end{figure} 

{\bf Two dimensions:}
In the square half-lattice the best estimate of $\mu_2$ (see equation \Ref{1b})
was used in equation \Ref{8}.  The top panel of figure \ref{f1} shows the data
for self-avoiding walks grafted to the hard wall (these are $10$-stars in our notation)
with entropic exponent $\gamma_1\equiv \gamma_{10}$. The middle curve shows
the confidence interval on the data as a shaded envelope.  There are odd-even 
parity effects in these plots, but they were dealt with by only plotting data for
even values of $n$. 

The best estimate
of $\gamma_1$ was obtained fixing its value to straighten the graph in figure \ref{f2}.  
An error bar on the best estimate was determined by varying the value of $\gamma_1$ to 
produce the other curves in the figure.  These curves are either convex or concave as 
the $y$-axis is approached, and this curving becomes apparent when $\gamma_1$ 
is changed by $\epsilon=0.00020$ in either direction.  Thus, we select as our error bar on 
$\gamma_1$ this value of $\epsilon$, giving our best estimate
\begin{equation}
\gamma_1 = 0.95325 \pm 0.00020 .
\end{equation}
The estimate for $\gamma_{11}$ was similarly obtained using the data of
grafted self-avoiding walks with their endpoints in the hard wall forming a surface
loop (these are $11$-stars in our notation).  The results of the analysis are shown
in the bottom panel of figure \ref{f2}.  Our best estimate is
\begin{equation}
\gamma_{11} = -0.1875 \pm 0.0020 .
\end{equation}
With these results one may test Barber's scaling relation.  Note that
\begin{equation}
2\gamma_1 - \gamma_{11} = 2.094 \pm 0.003 .
\label{17}
\end{equation}
The exact value is $\gamma+\nu = 43/32+3/4 = 67/32 = 2.09375$.
The absolute difference from the estimate above is
$0.00025$, a factor of $10$ smaller than the stated error bar of $0.003$.
This result is a strong numerical verification of Barber's scaling relation
in two dimensions.

\def\w{\textcolor{white}{$-$}}
\begin{table}[h!]
\caption{Half square lattice entropic exponents}
\begin{tabular}{ll@{}*{3}{l}l }
\;$\gamma_\C{G}$&
\quad Exact &
\multicolumn{2}{c}{Literature}& &
\w This work \cr  
\hline
\vspace{-2mm}
& & & & & \cr
\multirow{2}{*}{$\gamma_{1}$}         &\multirow{2}{*}{\w$0.953125$}\;  
                                                                &\w$0.945(5)$\,\cite{BGM78} 
                                                                &
                                                                &
                                                                &\multirow{2}{*}{\w$0.9533(2)$}  \cr
 & & \w$0.9551(3)$\,\cite{MC93} & & \cr
\vspace{-2mm}
& & & & & \cr
 $\gamma_{11}       $ &$-0.1875$\;  &$-0.19(3)$\,\cite{BGM78} 
                                                                     &\w   
                                                                     &\w   
                                                                     &$-0.188(2)$ \cr
\vspace{-2mm}
& & & & & \cr
 $\gamma_{20}       $ &\w$0.34375$\;  &\w
                                                                     &\w 
                                                                     &\w   
                                                                     &\w$0.344(1)$ \cr
\vspace{-3mm}
& & & & & \cr
 $\gamma_{21}      $ &$-0.796875$\;  &\w
                                                                     &\w
                                                                     &\w   
                                                                     &$-0.796(2)$ \cr
\vspace{-3mm}
& & & & & \cr
 $\gamma_{211}     $ &$-1.9375$\;  &\w
                                                                     &\w   
                                                                     &\w   
                                                                     &$-1.94(3)$ \cr
\vspace{-3mm}
& & & & & \cr
 $\gamma_{211}     $ &$-1.9375$\;  &\w
                                                                     &\w   
                                                                     &\w   
                                                                     &$-1.94(5)^{*}$ \cr
\vspace{-3mm}
& & & & & \cr
 $\gamma_{211}     $ &$-1.9375$\;  &\w
                                                                     &\w   
                                                                     &\w   
                                                                     &$-1.93(4)^{\dagger}$ \cr
\vspace{-2mm}
& & & & & \cr
 $\gamma_{30}       $ &$-0.828125$\;  &\w 
                                                                     &\w   
                                                                     &\w   
                                                                     &$-0.827(2)$ \cr
\vspace{-3mm}
& & & & & \cr
 $\gamma_{31}      $ &$-1.96875$\;  &\w
                                                                     &\w   
                                                                     &\w   
                                                                     &$-1.969(4)$ \cr
\vspace{-3mm}
& & & & & \cr
 $\gamma_{311}     $ &$-3.109375$\;  &\w 
                                                                     &\w   
                                                                     &\w   
                                                                     &$-3.11(1)$ \cr
\vspace{-3mm}
& & & & & \cr
 $\gamma_{3111}     $ &$-4.25$\;  &\w
                                                                     &\w   
                                                                     &\w   
                                                                     &$-4.25(5)^*$ \cr
\vspace{-3mm}
& & & & & \cr
\hline
\vspace{-3mm}
& & & & & \cr
\multicolumn{3}{l}{$*$ -- Calculated by equation \Ref{6}} \\
\multicolumn{3}{l}{$\dagger$ -- Calculated by equation \Ref{12}} \\
\end{tabular}
\label{t2}   
\end{table}

The remaining data for $2$-stars and $3$-stars were similarly analysed to obtain
estimates of the exponents $\gamma_{21^g}$ and $\gamma_{31^g}$.  The results
are shown in table \ref{t2}. Observe that there are no numerical estimates in the 
literature for these exponents.  These estimates are consistent, within error 
bars, with the (known) exact values in two dimensions.  This both 
confirms, on the one hand, that the exact results are correct, and on the other hand, 
that the numerical methods in this paper produce high quality estimates of the entropic 
exponents.   Testing equation \Ref{12} using our numerical values give
\begin{equation}
2\,\gamma_{21}-\gamma_{211} = 0.348 \pm 0.034
\label{18}
\end{equation}
and the exact value of $\gamma {-}1=11/32=0.34375$ is well inside the
stated error bar.  The estimate for $\gamma_{20}$ also verifies the
identity in equation \Ref{12x}. 

The exceptional case  is for $3111$-stars, namely $3$-stars
with their central node at the origin in the half square lattice and with 
each arm from the central node having its endpoint in the hard wall (see figure \ref{figure2}).
In this case the data were too sparse to analyse. In general very few $3111$-star 
conformations were encountered in our simulation because one arm will have
to be accommodated inside a surface loop formed by another and the hard wall (as
illustrated schematically in figure \ref{figure2}).  What data obtained were not inconsistent 
with the exact value $\gamma_{3111}=-4.25$.

A numerical estimate of $\gamma_{3111}$ can be obtained from the other three 
$3$-star exponents using equation \Ref{6}, namely
\begin{equation}
\gamma_{3111} = \gamma_{30}-3\,\gamma_{31}+3\,\gamma_{311}
= -4.25\pm0.05 ,
\end{equation}
where we used the results in table \ref{t2}, and added the error bars to
find a confidence interval. 

Overall, one can conclude that equation \Ref{4}, and the exact values 
of the entropic exponents in two dimensions, are supported by 
our numerical results.

\begin{table}[h!]
\caption{Half cubic lattice entropic exponents}
\begin{tabular}{ll@{}*{3}{l}l }
\;$\gamma_\C{G}$&
$\epsilon^1$-approx &
\multicolumn{2}{c}{Literature}& &
\w This work \cr  
\hline
\vspace{-2mm}
& & & & & \cr
\multirow{4}{*}{$\gamma_{1}$}         &\multirow{4}{*}{\w$0.625$}\;  
                                                                &\w$0.679(2)$\,\cite{HG94} 
                                                                &
                                                                &
                                                                &\multirow{2}{*}{\w$0.6785(8)$}  \cr
 & & \w$0.687(5)$\,\cite{LM88A,ML88A} & & \cr
 & & \w$0.6786(12)$\,\cite{G04}            & & 
                                                                &\multirow{2}{*}{\w$0.6776(10)^\ddagger$} \cr
 & & \w$0.677667(17)$\,\cite{CCG16}            & & &  \cr
\vspace{-2mm}
& & & & & \cr
\multirow{4}{*}{$\gamma_{11}$}       &\multirow{4}{*}{$-0.463$}\;  
                                                                &$-0.383(5)$\,\cite{HG94} 
                                                                &
                                                                &
                                                                &\multirow{2}{*}{$-0.389(3)$}  \cr
 & & $-0.38(2)$\,\cite{LM88A,ML88A} & & \cr
 & & $-0.390(2)$\,\cite{G04}               & & 
                                                                &\multirow{2}{*}{$-0.394(5)^\ddagger$} \cr
 & & $-0.389245(28)$\,\cite{CCG16}  & & &  \cr
\vspace{-2mm}
& & & & & \cr
\multirow{2}{*}{$\gamma_{20}$}       &\multirow{2}{*}{\w$0.125$}\;  
                                                                &\w$0.15698(34)^{+}$ \cite{D86,SBB11} & &
                                                                &\multirow{2}{*}{\w$0.154(3)$}  \cr
& & \w$0.15695300(95)^{+}$\,\cite{C17} & & \cr
\vspace{-2mm}
& & & & & \cr
\multirow{1}{*}{$\gamma_{21}$}       &\multirow{1}{*}{$-0.963$}\;  
                                                                & & &
                                                                &\multirow{1}{*}{$-0.918(8)$}  \cr
\multirow{1}{*}{$\gamma_{211}$}     &\multirow{1}{*}{$-2.050$}\;  
                                                                & & &
                                                                &\multirow{1}{*}{$-2.02(8)$}  \cr
\multirow{1}{*}{$\gamma_{211}$}     &\multirow{1}{*}{$-2.050$}\;  
                                                                & & &
                                                                &\multirow{1}{*}{$-1.99(2)^*$}  \cr
\multirow{1}{*}{$\gamma_{211}$}     &\multirow{1}{*}{$-2.050$}\;  
                                                                & & &
                                                                &\multirow{1}{*}{$-1.99(2)^\dagger$}  \cr
\vspace{-2mm}
& & & & & \cr
\multirow{1}{*}{$\gamma_{30}$}       &\multirow{1}{*}{$-0.500$}\;  
                                                                & & &
                                                                &\multirow{1}{*}{$-0.521(2)$}  \cr
\multirow{1}{*}{$\gamma_{31}$}       &\multirow{1}{*}{$-1.588$}\;  
                                                                & & &
                                                                &\multirow{1}{*}{$-1.59(2)$}  \cr
\multirow{1}{*}{$\gamma_{311}$}     &\multirow{1}{*}{$-2.675$}\;  
                                                                & & &
                                                                &\multirow{1}{*}{$-2.68(7)$}  \cr
\multirow{1}{*}{$\gamma_{3111}$}   &\multirow{1}{*}{$-3.763$}\;  
                                                                & & &
                                                                &\multirow{1}{*}{$-3.9(6)$}  \cr
\multirow{1}{*}{$\gamma_{3111}$}   &\multirow{1}{*}{$-3.763$}\;  
                                                                & & &
                                                                &\multirow{1}{*}{$-3.8(2)^*$}  \cr
\vspace{-2mm}
& & & & & \cr
\multirow{1}{*}{$\gamma_{40}$}       &\multirow{1}{*}{$-1.250$}\;  
                                                                & & &
                                                                &\multirow{1}{*}{$-1.325(4)$}  \cr
\multirow{1}{*}{$\gamma_{41}$}       &\multirow{1}{*}{$-2.338$}\;  
                                                                & & &
                                                                &\multirow{1}{*}{$-2.406(8)$}  \cr
\multirow{1}{*}{$\gamma_{411}$}     &\multirow{1}{*}{$-3.425$}\;  
                                                                & & &
                                                                &\multirow{1}{*}{$-3.48(4)$}  \cr
\multirow{1}{*}{$\gamma_{4111}$}   &\multirow{1}{*}{$-4.513$}\;  
                                                                & & &
                                                                &\multirow{1}{*}{$-4.6(2)$}  \cr
\multirow{1}{*}{$\gamma_{41111}$} &\multirow{1}{*}{$-5.600$}\;  
                                                                & & &
                                                                &\multirow{1}{*}{$-5.8(1.1)^*$}  \cr
\vspace{-2mm}
& & & & & \cr
\multirow{1}{*}{$\gamma_{50}$}       &\multirow{1}{*}{$-2.125$}\;  
                                                                & & &
                                                                &\multirow{1}{*}{$-2.243(4)$}  \cr
\multirow{1}{*}{$\gamma_{51}$}       &\multirow{1}{*}{$-3.213$}\;  
                                                                & & &
                                                                &\multirow{1}{*}{$-3.318(7)$}  \cr
\multirow{1}{*}{$\gamma_{511}$}     &\multirow{1}{*}{$-4.300$}\;  
                                                                & & &
                                                                &\multirow{1}{*}{$-4.41(4)$}  \cr
\multirow{1}{*}{$\gamma_{5111}$}     &\multirow{1}{*}{$-5.388$}\;  
                                                                & & &
                                                                &\multirow{1}{*}{$-5.5(2)$}  \cr
\vspace{-3mm}
& & & & & \cr
\hline
\vspace{-3mm}
& & & & & \cr
\multicolumn{6}{l}{$*$ -- Calculated by equation \Ref{6}} \\
\multicolumn{6}{l}{$\dagger$ -- Calculated by equation \Ref{12}} \\
\multicolumn{6}{l}{$\ddagger$ -- Panneton random number generator \cite{PLM06}} \\
\multicolumn{6}{l}{${+}$ -- Estimations using equation \Ref{12x}} \\
\end{tabular}
\label{t3}   
\end{table}

{\bf Three dimensions:}
In the cubic lattice corrections to scaling are dominated by the confluent
correction term which is of the form $C/n^\Delta$ (see equation \Ref{9}).
There is a competing, faster decaying, analytic correction $B/n$, or even higher
order confluent corrections, which may impact an analysis using 
equation \Ref{8}, in particular at small values of $n$.  Our approach here
is based on the methods developed in reference \cite{G04}, and we
will be plotting the left hand of equation \Ref{8} against $n^{-\Delta}$
where $\Delta$ is fixed at its best estimate $\Delta=0.528(8)$
\cite{CD16}.  Unlike in the square lattice, the confluent correction decays
slowly, and competing higher order corrections may impact the analysis
at small values of $n$.  Therefore, the aim here is to find a linear plot
at large values of $n$, discarding data at small $n$.
In addition, there are, like in the square lattice, odd-even parity effects in 
the data, and we dealt with these by only plotting data for even values of $n$. 

\begin{figure}[h!]
\includegraphics[width=0.475\textwidth]{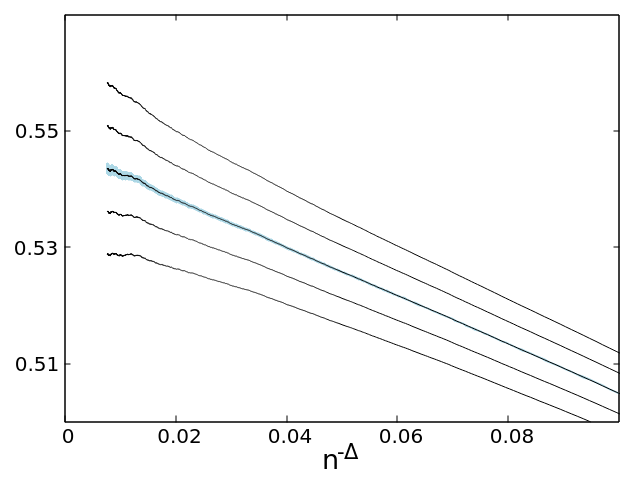} \\
\vspace{3mm}
\includegraphics[width=0.475\textwidth]{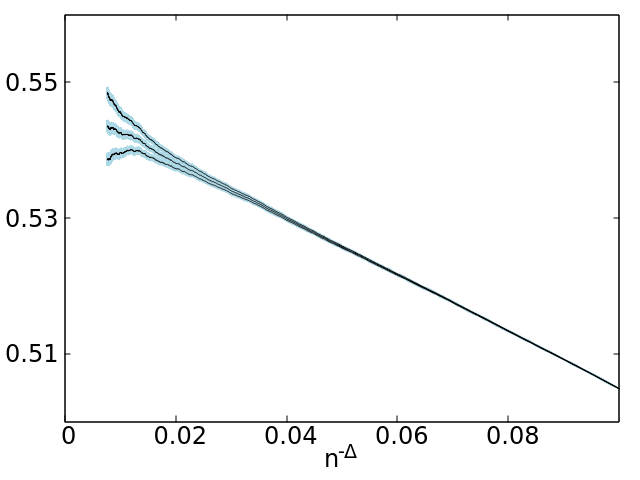}
\caption{\textit{Determining the best value of $\gamma_1$.  In
the top panel equation \Ref{8} was plotted with $\Delta=0.528$
and $\mu_3=4.684039931-\delta$, where $\delta=2.3\times10^{-6}$.
The effect of minor changes in the plots due to changes in 
$\mu_3$ are shown in the bottom panel.  The top curve is for
$\mu_3$ equal to its best value in equation \Ref{1c}, the middle
curve for $\mu_3-\delta$, and the bottom curve for $\mu_3-2\delta$.}}
\label{f4}  
\end{figure}

Plotting equation \Ref{8} to determine $\gamma_1$ gave graphs which
were typically first concave (at small values of $n$) and then turning convex
at large values of $n$.  A linear graph can be obtained by making minor
changes in the estimate of $\mu_3$ and this gives the best estimate
\begin{equation}
\gamma_1 = 0.6745\pm 0.0008 . 
\label{14}
\end{equation}
This estimate was obtained by determining that value of $\gamma_1$ that
straightens the graph when equation \Ref{8} was plotted using the data.  In this case
its final value was insensitive to small changes in the value of $\mu_3$
from its best estimate in equation \Ref{1c}, and also with small changes 
in $\Delta$ within its error bars.  However, it was not possible to find a straight 
graph of the data without changing the value of $\mu_3$ in minor ways from 
that in equation \Ref{1c}.  Following the approach in reference \cite{G04} the effects
of these small changes in $\mu_3$ are shown in the bottom panel of figure \ref{f4}, 
where our data are plotted with $\gamma_1$ fixed at its best value in equation 
\Ref{14}, but with the growth constant fixed at $\mu_3{-}k\,\delta$ where 
$k=0,1,2$ and $\delta=2.3\times 10^{-6}$.  If $k=0$, then the curvature 
is upwards as the $y$-axis is approached.  The bottom downwards curvative is
seen for $k=2$, while the best fit is for $k=1$.  

These results do \textit{not}
imply that the estimate in equation \Ref{1c} is suspect, but instead expose
limitations in the data in this paper -- if the purpose was to estimate $\mu_3$
from the data obtained here, then one would at best expect to do this to an 
accuracy of $O(2.3\times 10^{-6})$.  In addition, changing $\mu_3$ from its
best value introduces an extra degree of freedom in the analysis, and may give
biassed estimates of the exponents.  In order to avoid this possibility, we fixed
the value of $\mu_3$ at its best known estimate, and then proceeded with
curve fitting while discarding data an small values of $n$.

\begin{figure}[h!]
\includegraphics[width=0.475\textwidth]{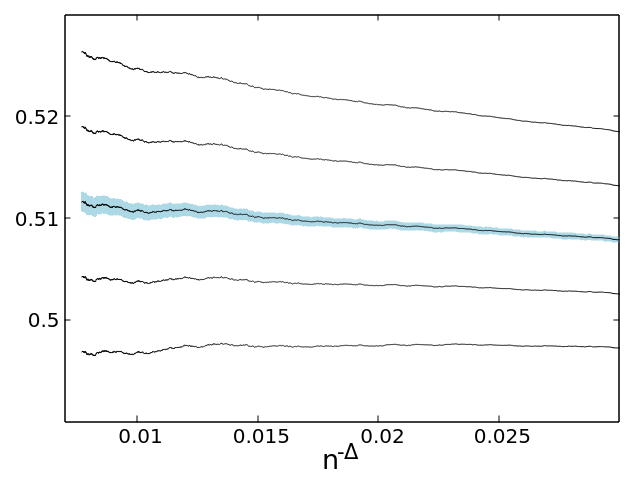} 
\caption{\textit{Determining the best value of $\gamma_{1}$.
The middle curve is a plot of equation \Ref{8} as a function of 
$1/n^\Delta \in [0.007,0.03]$ with $\Delta=0.528$ and $\mu_3=4.684039931$.  
This gives the estimate $\gamma_{1} = 0.6785(8)$.  The top two graphs 
are plotted using $\gamma_{1}+k\epsilon$  with $k=1,2$ where 
$\epsilon=0.0008$, while the bottom two are plots with $k=-1,-2$.}}
\label{f4-b}  
\end{figure}

The analysis giving the estimate in equation \Ref{14} relies almost exclusively
on data with $0.02 \leq n^{-\Delta} \leq 0.10$ (as seen in the bottom panel
of figure \ref{f4}).  This corresponds to $78 \leq n \leq 1650$, while data with
$n>1650$ are compensated by the small change in the value of $\mu_3$.
This, however, cannot be the best way of extracting a good estimate of $\gamma_1$,
and more care is needed.  In particular, one should rely on large values of $n$
in the analysis, since corrections to scaling are reduced with increasing $n$.
In addition, changes in the value of $\mu_3$ introduces an additional 
degree of freedom in the analysis, and it primarily affects data at large $n$.
Thus, the analysis was repeated, but without changes in the value of $\mu_3$, and 
discarding data with $n\leq 766$. This gives the results in figure \ref{f4-b}.  
The middle curve corresponds to the best estimate of $\gamma_1$: 
\begin{equation}
\gamma_1 = 0.6785 \pm 0.0008 .
\label{141}
\end{equation}
The two top curves are convex, while the two bottom curves are concave,
and give the estimated error bar above. Since this estimate is based on data for 
larger values of $n$, it is taken as the best estimate of this exponent.

We have listed our best estimate for $\gamma_1=0.6785(8)$ in table \ref{t3},
where we also compare it to earlier estimates in the literature.  The best 
available estimate is in reference \cite{CCG16}.  This estimate excludes 
our best estimate from its (very small) error bar.  Conversely, that estimate
is well within the error bar in equation \Ref{141}.  Sources of a systematic
error in our estimate may be due to the choice of random number 
generator (in this paper we used the mersenne twistor for 64 bit architecture 
\cite{MN98}), or due to limitations in carrying significant digits along in the 
simulation (we used long double ($80$-bit) precision in the C programming language).
We also used gnuplot \cite{gnuplot} to analyse the data, and it also has finite
precision.  We avoided large numbers in our simulation by only storing
the ratio $s_n^{(f)}(g)/\mu_3^n$ in our programs and data files.  Repeated
division by $\mu_3$ during the simulation may also introduce rounding errors 
which accumulate during the simulation. 

\begin{figure}[h!]
\includegraphics[width=0.475\textwidth]{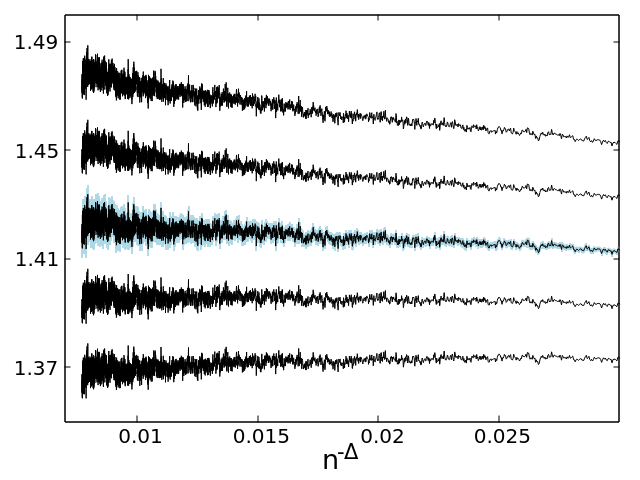} 
\caption{\textit{Determining the best value of $\gamma_{11}$.
The middle curve is a plot of equation \Ref{8} as a function of $1/n^\Delta$
with $\Delta=0.528$ and $\mu_3=4.684039931$.  This gives the estimate 
$\gamma_{11}= -0.3893$. The top two graphs are plotted using 
$\gamma_{11}+k\epsilon$ with $k=1,2$ where $\epsilon=0.0030$, while 
the bottom two are plots with $k=-1,-2$.}}
\label{f6}  
\end{figure}

The value of the surface loop exponent $\gamma_{11}$ was similarly estimated
(figure \ref{f6}).  The best estimate consistent with our data is
\begin{equation}
\gamma_{11} = -0.389 \pm 0.003 .
\label{16}
\end{equation}
Barber's scaling relation can be tested in three dimensions by the results
in equations \Ref{14} and \Ref{16}.  The best available estimate of the
numerical estimate of the metric exponent of self-avoiding walks is 
$\nu = 0.58759700(40)$ \cite{C10,CD16} and of the entropic exponent 
$\gamma=1.156 953 00(95)$ \cite{C17}. Adding these gives
$\gamma{+}\nu = 1.7445500(14)$.  Our results give
\begin{equation}
2\,\gamma_{1}-\gamma_{11} = 1.746(5) .
\label{22}
\end{equation}
This result includes the sum $\gamma+\nu$ inside its error bar, and so is
consistent with the Barber scaling relation in three dimensions. 

These results were retested by performing simulations using an alternative random
number generator (the Panneton generator \cite{PLM06}).  The results
are shown in table \ref{t3} where $\gamma_1 = 0.6776(10)$ and 
$\gamma_{11}=-0.394(5)$. This gives $2\gamma_1-\gamma_{11}
=1.749(7)$, again consistent with the Barber scaling relation and with
equation \Ref{22}.
%
%

Plotting our data for grafted $f$-stars with $2\leq f\leq5$ produced graphs which 
do not straighten at the best value of $\gamma_{f1^g}$.  Instead, the locus 
of the data points were typically concave at small values of $n$, even as it straightens
as $n$ increases.  This again suggests that higher order corrections to scaling are 
complicating the analysis.  Since the slowest decaying correction is $C/n^\Delta$, and 
it becomes dominant as $n$ is increased,  the exponents were estimated
by focussing on the largest values of $n$ as before.  That is, by using equation \Ref{8} the
exponent is estimated by setting it to straighten the curve at the largest values of
$n$, even if there is a remaining curvature seen at the smallest values of $n$.

\begin{figure}[h!]
\includegraphics[width=0.475\textwidth]{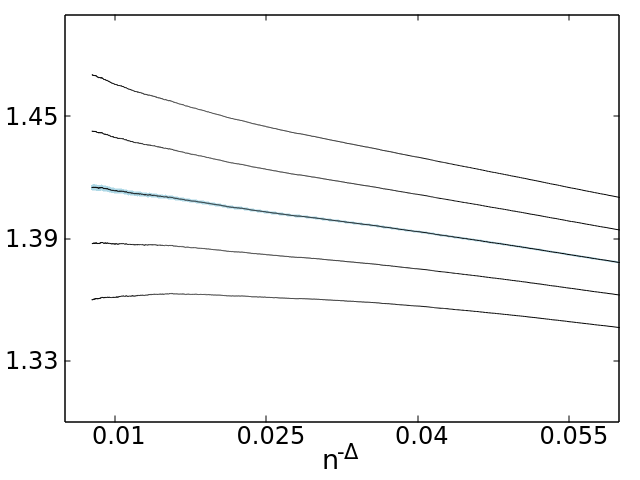} 
\caption{\textit{Estimating $\gamma_{20}$ by plotting the left hand side 
of equation \Ref{8} as a function of $n^{-\Delta}$ for $206\leq n \leq 10000$.
The middle graph corresponds to the best esimate $\gamma_{20}=0.154\pm0.003$,
while the top two curves, and the bottom two curves, are used to determine
the confidence interval.}}
\label{f8}  
\end{figure}

In figure \ref{f8} the result for $\gamma_{20}$ is shown.  These graphs were
obtained by using the best estimate obtained from our data and give
\begin{equation}
\gamma_{20} = 0.154 \pm 0.003 .
\label{22a}
\end{equation}
The best estimate of $\gamma_{20}$ in the literature is obtained by noting
from equation \Ref{12x} that that $\gamma_{20}=\gamma{-}1$ and 
using the best estimate $\gamma=1.156 953 00(95)$ \cite{C17}.
This shows that  $\gamma_{20}= 0.156 953 00(95)$ and this is well within the stated 
error bar of the estimate in equation \Ref{22a}.  Conversely, these results also
support the identity in equation \Ref{12x} in three dimensions.

\begin{figure}[h!]
\includegraphics[width=0.475\textwidth]{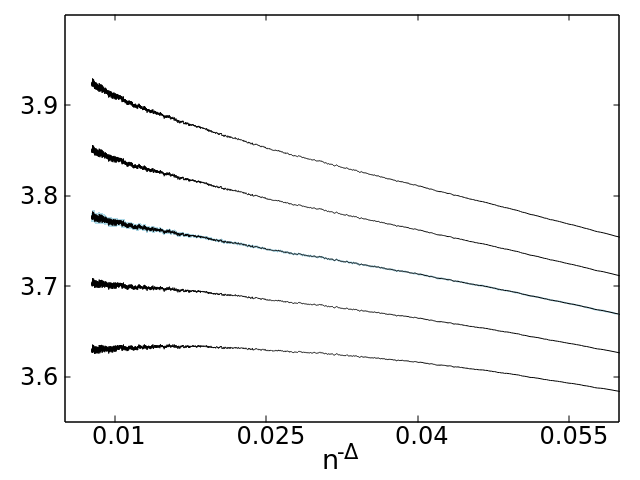} 
\caption{\textit{Plotting equation \Ref{8} against $n^{-\Delta}$ for $573\leq n
\leq 10000$ to determine $\gamma_{21}$.}}
\label{f9}  
\end{figure}

The analysis for $\gamma_{21}$ is shown in figure \ref{f9}.  These graphs are
for $573 \leq n \leq 10000$. Observe that there remains a minor concavity 
in the middle curve at the largest values of $n^{-\Delta}$ but that the curves
straigthen as $n^{-\Delta}$ decreases when $n$ approaches $n=10,000$.  The 
top two curves are convex, and the bottom two curves are concave.  This gives 
the best value of $\gamma_{21}$:
\begin{equation}
\gamma_{21} = -0.918\pm 0.008 .
\label{23}
\end{equation}

\begin{figure}[t!]
\includegraphics[width=0.475\textwidth]{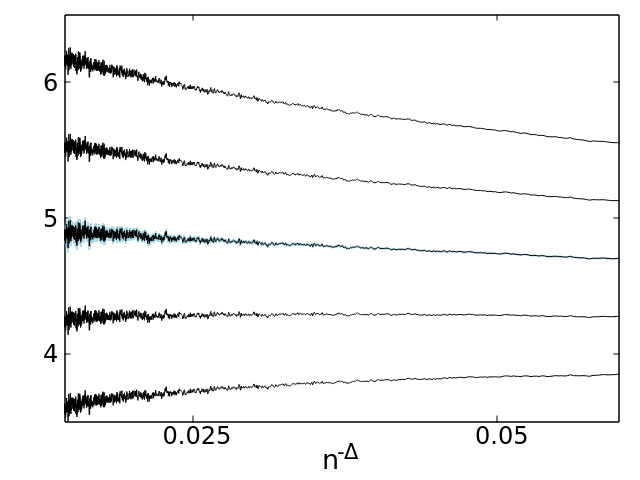} 
\caption{\textit{Plotting equation \Ref{8} against $n^{-\Delta}$ for $37\leq n
\leq 2946$ to determine $\gamma_{211}$.}}
\label{f10}  
\end{figure}

The estimates of $\gamma_{20}$ and $\gamma_{21}$ in equations
\Ref{22} and \Ref{23} can be used 
to predict $\gamma_{211}$ using equation \Ref{6}.  This gives
\begin{equation}
\gamma_{211} = 2\,\gamma_{21}-\gamma_{20} = -1.99 \pm 0.02 .
\label{24}
\end{equation}
Determining $g_{211}$ directly from the data is complicated by poor
sampling at large $n$.  Examination of the data shows reasonable
sampling for $n\leq 2500$, and poor sampling for $n\geq 3000$.
Plotting equation \Ref{8} for $37\leq n \leq  2846$ gives figure \ref{f10}
which unambiguously gives the estimate
\begin{equation}
\gamma_{211} = -2.02 \pm 0.08 
\label{26}
\end{equation}
with a conservatively determined error bar (there is significant
curvature present in the second and fourth curves in figure \ref{f10}).
This result is consistent with the estimated value in equation \Ref{24}.
Using this result with the estimate of $\gamma_{21}$ gives
$2\gamma_{21}-\gamma_{211} = 0.18(10)$.  Within its large error bar,
this result is consistent with $\gamma=0.156 953 00(95)$ \cite{C17}
as shown by equation \Ref{12}.

The estimates for grafted $2$-star exponents are listed in table \ref{t3}.

Data for grafted $3$-, $4$- and $5$-stars were similarly analysed and the results
appear in table \ref{t3}.  In the case of grafted $3$-stars, the estimates are
\begin{align}
\gamma_{30} &= -0.521 \pm 0.002, \nonumber \\
\gamma_{31} &= -1.59 \pm 0.02, 
\label{27} \\
\gamma_{311} &= -2.68 \pm 0.07, \nonumber \\
\gamma_{3111} &= -3.9 \pm 0.6 . \nonumber
\label{27}
\end{align}
The estimate for $\gamma_{3111}$ is based on data for $206 \leq n \leq 1080$.
These results are consistent with equation \Ref{6}. Using equation \Ref{6} 
and the estimates for $\gamma_{30}$, $\gamma_{31}$ and $\gamma_{311}$
gives a better estimate of $\gamma_{3111}$ instead:
\begin{equation}
\gamma_{3111} = \gamma_{30}-3\,\gamma_{31}+3\,\gamma_{311}
= -3.8 \pm 0.3 ,
\label{28}
\end{equation}
and this result is still consistent with the estimate of $\gamma_{3111}$ in
equation \Ref{27}.

The data for grafted $4$s-stars give
\begin{align}
\gamma_{40} &= -1.325 \pm 0.004, \nonumber \\
\gamma_{41} &= -2.406 \pm 0.008, \\
\gamma_{411} &= -3.48 \pm 0.04, \nonumber \\
\gamma_{4111} &= -4.6 \pm 0.2. \nonumber 
\end{align}
The estimate for $\gamma_{4111}$ is based on $206 \leq n \leq 1650$.
By equation \Ref{6},
\begin{equation}
\gamma_{41111} = - \gamma_{40}+4\,\gamma_{41}-6\,\gamma_{411}
+4\,\gamma_{4111} = -5.8 \pm 1.1 .
\end{equation}
Finally, for grafted $5$-stars
\begin{align}
\gamma_{50} &= -2.251 \pm 0.003, \nonumber \\
\gamma_{51} &= -3.333 \pm 0.007, \\
\gamma_{511} &= -4.41 \pm 0.04, \nonumber \\
\gamma_{5111} &= -5.5 \pm 0.2. \nonumber 
\end{align}
Our sampling of $51111$- and $511111$-stars were too poor to allow
estimates of the entropic exponents.

\section{Conclusions}
 
The purpose of this paper was to estimate the entropic exponents of 
half-space grafted $f$-stars, and to numerically verify some relations involving these
exponents.  Our results are shown in tables \ref{t2} and \ref{t3},
Barber's scaling relation is tested in equations \Ref{17} and
\Ref{22}, and equations \Ref{12x} and \Ref{12} were tested in equations
\Ref{18} and \Ref{22a}.  The relation in equation \Ref{12} was similarly
tested for grafted $2$-stars and $3$-stars in the half cubic lattice
(equations \Ref{24} and \Ref{26}, and \Ref{27} and \Ref{28}). In all
respects the general framework using vertex exponents $\sigma_f$
and surface vertex exponents $\sigma_f^\prime$ in equation \Ref{3}
is strongly supported by the numerical results here.  
 
The results in two dimensions are consistent to good accurately with the 
exact (conformal invariance) values of the exponents.  This not only
provides strong evidence supporting the theoretical analysis of the surface
entropic exponents for uniform branched networks in two dimensions
in references \cite{D86,D89C,C87,N84}, but also shows that the numerical methods 
used in this paper (and in references \cite{HNG04,GG94,G04,HG94}) are sound.
This enhances confidence in the cubic lattice results shown here, which cannot
be verified against a list of exact values.  On the contrary, few of the 
surface exponents of grafted lattice stars in three dimensions have been calculated 
before (as can be seen in table \ref{t3}), apart from the $O(\epsilon)$-expansion
estimates which give good, but not excellent, agreement with the numerical
estimates obtained in this paper.

 \section*{Acknowledgements} EJJvR acknowledges financial support 
from NSERC (Canada) in the form of Discovery Grant RGPIN-2019-06303
and is in debt to N Clisby for useful feedback on an earlier version of the 
manuscript.

\bibliographystyle{unsrt}
\bibliography{halfspacestars.bib}

\end{document}